\documentclass[10pt,conference]{IEEEtran}

\usepackage{booktabs}
\usepackage{hyperref}
\usepackage{amsmath}
\usepackage{verbatim}
\usepackage{enumerate}
\usepackage{siunitx}
\usepackage[shortlabels]{enumitem}
\usepackage{fancybox}
\usepackage{url}
\usepackage{flushend}
\usepackage{graphicx}
\usepackage{setspace}
\usepackage[roman]{parnotes}
\usepackage{color}
\usepackage{colortbl}
\usepackage{listings}
\usepackage{tabulary}
\usepackage{xcolor}
\usepackage{framed}
\usepackage{subfigure}
\usepackage{xspace}
\usepackage[symbol]{footmisc}

\newcounter{observation}
\newcommand{\observation}[1]{\refstepcounter{observation}
	\begin{center}
		\framebox{
			\begin{minipage}{0.93\columnwidth}
				{} \textit{#1}
			\end{minipage}
		}
	\end{center}
}

\newcommand{\plugin}{\textsc{PyNose}\xspace}
\newcommand{\miner}{\textsc{PythonChangeMiner}\xspace}
\newcommand{\smell}{\textit{Suboptimal Assert}\xspace}
\newcommand{\main}{\textit{Primary}\xspace}
\newcommand{\control}{\textit{Secondary}\xspace}

\makeatother

\setlist{nosep}
\setlist[itemize]{leftmargin=2.3em}
\setlist[enumerate]{leftmargin=2.3em}
\makeatletter
\newcommand{\linebreakand}{%
  \end{@IEEEauthorhalign}
  \hfill\mbox{}\par
  \mbox{}\hfill\begin{@IEEEauthorhalign}
}
\makeatother

\begin{document}

\title{\plugin: A Test Smell Detector For Python}

\author{
\IEEEauthorblockN{Tongjie Wang*}
\IEEEauthorblockA{\textit{University of California, Irvine}\\
Irvine, CA, United States \\
tongjiew@uci.edu}

\and

\IEEEauthorblockN{Yaroslav Golubev*}
\IEEEauthorblockA{\textit{JetBrains Research}\\
Saint Petersburg, Russia \\
yaroslav.golubev@jetbrains.com}

\and

\IEEEauthorblockN{Oleg Smirnov}
\IEEEauthorblockA{\textit{JetBrains Research}\\
\textit{Saint Petersburg State University}\\
Saint Petersburg, Russia \\
oleg.smirnov@jetbrains.com}

\linebreakand

\IEEEauthorblockN{Jiawei Li}
\IEEEauthorblockA{\textit{University of California, Irvine}\\
Irvine, CA, United States \\
jiawl28@uci.edu}

\and

\IEEEauthorblockN{Timofey Bryksin}
\IEEEauthorblockA{\textit{JetBrains Research}\\
\textit{Saint Petersburg State University}\\
Saint Petersburg, Russia \\
timofey.bryksin@jetbrains.com}

\and

\IEEEauthorblockN{Iftekhar Ahmed}
\IEEEauthorblockA{\textit{University of California, Irvine}\\
Irvine, CA, United States \\
iftekha@uci.edu}
}

\maketitle

\begin{abstract}

Similarly to production code, code smells also occur in test code, where they are called \textit{test smells}. Test smells have a detrimental effect not only on test code but also on the production code that is being tested. To date, the majority of the research on test smells has been focusing on programming languages such as Java and Scala. However, there are no available automated tools to support the identification of test smells for Python, despite its rapid growth in popularity in recent years. In this paper, we strive to extend the research to Python, build a tool for detecting test smells in this language, and conduct an empirical analysis of test smells in Python projects.

We started by gathering a list of test smells from existing research and selecting test smells that can be considered language-agnostic or have similar functionality in Python's standard \textit{Unittest} framework. 
In total, we identified 17 diverse test smells. Additionally, we searched for Python-specific test smells by mining frequent code change patterns that can be considered as either fixing or introducing test smells. Based on these changes, we proposed our own test smell called \smell. To detect all these test smells, we developed a tool called \plugin in the form of a plugin to PyCharm, a popular Python IDE.
Finally, we conducted a large-scale empirical investigation aimed at analyzing the prevalence of test smells in Python code. Our results show that 98\% of the projects and 84\% of the test suites in the studied dataset contain at least one test smell. Our proposed \smell smell was detected in as much as 70.6\% of the projects, making it a valuable addition to the list.    

\end{abstract}


\section{Introduction}

\renewcommand*{\thefootnote}{\fnsymbol{footnote}}
\footnotetext[1]{The ﬁrst two authors contributed equally to this work.}
\renewcommand*{\thefootnote}{\arabic{footnote}}

\textit{Code smells} were introduced to identify potential maintainability issues in software systems~\cite{fowler1999refactoring}, however, later they have been used as a measure of design quality of software projects~\cite{deligiannis2003empirical,li2007empirical,oliva2013can}. Researchers found that code smells are associated with bugs~\cite{li2007empirical,olbrich2009evolution}, fault-proneness~\cite{hall2014some,zazworka2011investigating}, and maintainability issues in the code base~\cite{fowler1999refactoring}. 
While investigating the underlying reasons for introducing code smells,
researchers attributed various factors to this, including developers struggling with deadlines~\cite{tufano2017and} or not caring about the impact of the applied design choices~\cite{fowler1999refactoring}.

Similarly to production code, test code can also have code smells, in which case they are called \textit{test smells}. Van Deursen et al.~\cite{van2001refactoring} defined test smells as being caused by poor design choices (similarly to regular code smells) when developing test cases.\footnote{ To avoid the ambiguity that exists in testing terminology between languages and frameworks, in this paper, we will always refer to individual tests or test methods as \textit{test cases} and to classes that group them as \textit{test suites}.} Just like the code smells, test smells make the impacted test code harder to maintain and comprehend~\cite{bavota2012empirical}. Moreover, recent studies have shown that test smells also impact the quality of production code~\cite{spadini2018relation}.

Since test smells have a negative impact on the quality of production code, it is of great interest and importance to study and detect them. To date, the majority of the research on test smells has been focusing on statically typed languages like Java and Scala~\cite{bavota2012empirical,spadini2018relation,bavota2015test,tufano2016empirical,greiler2013automated,de2019assessing}. However, in recent years, Python has been growing in popularity due to being the primary language used in Data Science and Machine Learning in particular~\cite{raschka2020machine,sarkar2018practical}. Furthermore, despite the empirical evidence against test smells, developers tend not to be aware of the smells that exist in their tests~\cite{tufano2016empirical}, and the lack of efficient tools can be one of the reasons for it. To the best of our knowledge, there are no works that study the existence and prevalence of test smells in Python code, and no tools exist that specifically aim at identifying test smells in this language.

In this paper, we aim to fill these gaps by curating a list of possible test smells for Python, a tool for their detection, and an empirical study of their pervasiveness in Python code. We started by conducting a small-scale mapping study to find different test smells studied in the literature and selecting test smells that can be considered language-agnostic or have analogous functionality in Python's standard \textit{Unittest} framework. In total, we identified 17 diverse test smells. These test smells were all adopted from other papers dedicated to different programming languages, but it is natural to assume that Python has its own specific test smells. To discover them, we used a tool called \miner~\cite{golubev2021changes} to search for frequent change patterns in test suites.
We manually evaluated 159 patterns that occur in at least three different projects and identified 32 possible changes that are related to \textit{assert} functions in \textit{Unittest} and are aimed at making the tests more specific and simplify the understanding of the testing logic. We bundled the less specific versions of these assertions together into a single \smell test smell. Thus, a total of 18 smells were identified for Python.

We developed \plugin, a plugin for PyCharm~\cite{pycharm} that is able to detect these smells in the Python code.
Using the tool, we performed an empirical study on the prevalence of test smells in 248 Python projects. Our results indicate that test smells are indeed common in Python test code, with 98\% of projects and 84\% of test suites having at least one test smell.

Overall, our contributions are as follows:

\begin{itemize}
    \item We conducted a small-scale mapping study and compiled a list of test smells that are applicable to Python. 
    \item We identified a new Python-specific test smell by analyzing Python test code changes.
    \item We developed a tool called \plugin as a plugin for PyCharm that can detect test smells from Python projects that use the standard \textit{Unittest} framework. \plugin is available for researchers and practitioners on GitHub: \url{https://github.com/JetBrains-Research/PyNose}.
    \item We report the findings pertaining to the pervasiveness of test smells from an empirical study conducted on 248 Python projects.
\end{itemize}

The rest of the paper is organized as follows. In Section~\ref{sec:related}, we discuss the existing works in the field of test smells detection and analysis. Section~\ref{sec:selecting} describes the choice of test smells for Python and the search for Python-specific test smells. In Section~\ref{sec:tool}, we describe the development of \plugin and its evaluation, and in Section~\ref{sec:study}, we describe the empirical study that we conducted using the tool, as well as its results. In Section~\ref{sec:threats}, we discuss threats to the validity of our study, and, finally, in Section~\ref{sec:conclusion}, we conclude our paper and discuss possible future work.


\section{Related Work}
\label{sec:related}

\begin{figure*}
\centering
\includegraphics[width=\textwidth]{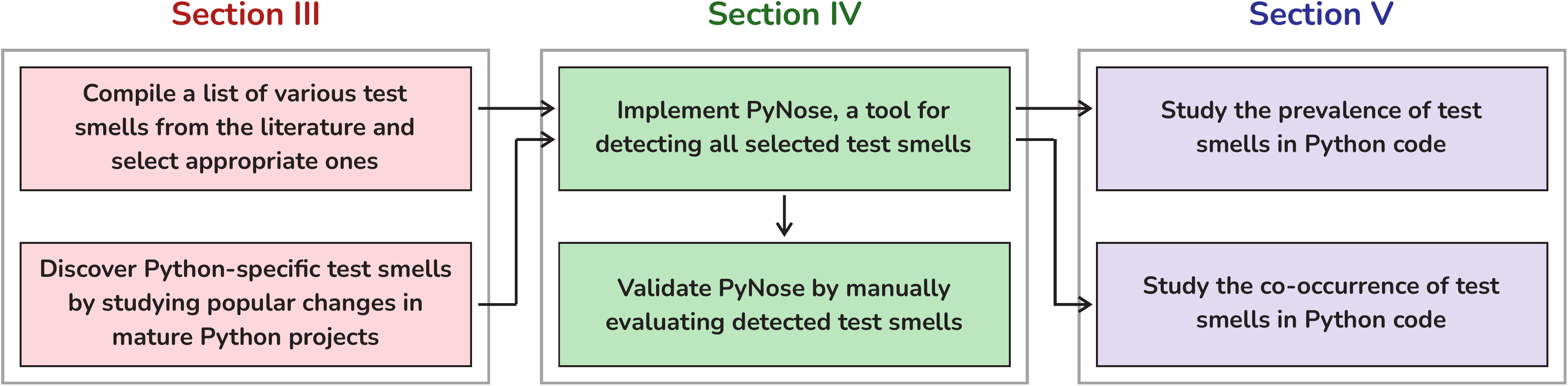}
\caption{The overall pipeline of the study.}
\label{fig:pipeline}
\end{figure*}

Similarly to production code, test code should be designed following proper established programming practices~\cite{reitz2016hitchhiker}. Van Deursen et al.~\cite{van2001refactoring} defined the term \textit{test smells} as code smells that are caused by poor design choices when developing test cases and also defined a catalog of 11 test smells. Later, several researchers extended this catalog~\cite{greiler2013automated,meszaros2007xunit,van2006characterizing,breugelmans2008testq}. While the majority of the research focused on test smells occurring in Java, several researchers investigated other languages and domains. For example, Bleser et al. investigated test smells in Scala~\cite{de2019assessing, de2019socrates}, while Peruma et al.~\cite{peruma2019distribution} explored unit tests in mobile applications and identified several new test smells.

Researchers have also been investigating the negative impacts of test smells on software development~\cite{bavota2012empirical,spadini2018relation,bavota2015test,tufano2016empirical,virginio2019influence}. By conducting two empirical studies, Bavota et al.~\cite{bavota2012empirical,bavota2015test} showed that test smells are widely spread throughout software systems, and most test smells have a strong negative impact on the comprehensibility of test suites and production code. Spadini et al.~\cite{spadini2018relation} investigated the relationship between the presence of test smells and the change- and defect-proneness of test code, as well as the defect-proneness of the tested production code. They found that some test smells are more change-prone than others, and they also found that production code tested by smelly tests is comparatively more defect-prone. Tufano et al.~\cite{tufano2016empirical} found that test smells are usually introduced when the corresponding test code is committed to the repository for the first time, and they tend to remain in a system for a long time. Virgínio et al.~\cite{virginio2019influence} investigated correlations between test coverage and test smells, and found that test smells influence code coverage.

Investigating ways for an automated detection of test smells has also received attention from the research community. Van Rompaey et al.~\cite{van2007detection} proposed a set of metrics defined in terms of unit test concepts and compared the proposed detection techniques effectiveness with human review. Greiler et al.~\cite{greiler2013automated} analyzed the relationship between the development of a test fixture and possible test smells within it. They also designed a static analysis tool to identify fixture-related test smells and evaluated them by discovering test smells in three industrial projects. Palomba et al.~\cite{palomba2018automatic} developed an automated textual-based approach for detecting several types of test smells. Compared with the \textit{code metrics}-based techniques proposed by Greiler et al. and Van Rompaey et al., the textual-based technique proved to be more effective in detecting certain test smells. Peruma et al.~\cite{peruma2019distribution, peruma2020tsdetect} recently developed a tool called \textsc{tsDetect} capable of detecting 19 test smells in Java.

More recently, researchers have been investigating ways to help testers refactor test smells. Lambiase et al.~\cite{lambiase2020just} presented an IntelliJ-based plugin that enables an automated identification and refactoring of test smells using IntelliJ Platform's APIs. Santana et al.~\cite{santana2020raide} proposed another tool that can be used in an IDE, providing testers with an environment for automated detection of lines of code affected by test smells, as well as a semi-automated refactoring for Java projects. Virgínio et al.~\cite{virginio2020jnose} presented a tool designed to analyze test suite quality in terms of test smells. Their tool is the first one that relies on both code coverage and the presence of test smells to measure the quality of tests. 

Overall, the majority of the mentioned research has been focusing on Java. However, in recent years, Python has been growing more popular because of its important role in Data Science and Machine Learning in particular~\cite{raschka2020machine,sarkar2018practical}. To the best of our knowledge, no tools exist that specifically aim at identifying Python test smells. \plugin addresses this gap. Furthermore, there is no large-scale analysis regarding the prevalence of test smells in Python code, and our study is the first towards filling this gap in research.
\section{Selecting Test Smells}
\label{sec:selecting}

The goal of our study is to build a tool that can identify test smells in Python code as well as to assess to what extent test smells are prevalent in Python test suites. The general pipeline of our study is demonstrated in Figure~\ref{fig:pipeline}. In Section~\ref{sec:selecting}, we curate the list of appropriate test smells by conducting a systematic mapping study (Section~\ref{sec:literature}) and then augmenting the list by identifying Python-specific test smells (Section~\ref{sec:python-specific}).

\subsection{Systematic mapping study of test smells}
\label{sec:literature}
As a first step, we conducted a small-scale systematic mapping study on test smells to curate a list of test smells discussed in the literature. According to Kitchenham et al.~\cite{kitchenham2015evidence}, the goal of the mapping study is to survey the available knowledge about a topic. 

\textit{Search Question.} Our search question was phrased as follows:
\textit{What test smells have been studied in literature to date?} 

\textit{Search Keywords.} To determine the optimal set of search keywords, we conducted a pilot search on two well-known digital libraries, IEEE and ACM. This process was intended to identify relevant words utilized in test smell publications. We conducted our query only on the title and abstract of the publication to avoid false positives. 
The finalized search string is presented below.

\observation{\textbf{Title}: (``test smell'' OR ``test smells'') AND \textbf{Abstract}: (``test smell'' OR ``test smells'').}

\textit{Data Source.} To discover relevant publications, we used three of the most popular online paper search engines: ACM Digital Library, IEEE Xplore, and Scopus.

\textit{Search Period.} To obtain as many related works as possible, we queried all related studies before 2020.
This resulted in a list of papers that were published between 2006 to 2020.

\textit{Initial Results.}
Our initial search of the three digital libraries resulted in 54 publications. To narrow down the search results, next, we filtered out publications that were not part of our inclusion criteria. A summary of the inclusion and exclusion criteria used to filter the retrieved literature is shown in Table~\ref{tab:filter_criteria}.
The filtering process helped us to reduce the number of studies significantly, however, this may have resulted in leaving out some relevant studies. Thus, we conducted backward snowballing~\cite{wohlin2014guidelines} (\textit{i.e.,} looking for additional studies in the reference lists of the selected studies, as suggested by Keele et al.~\cite{keele2007guidelines}). In our work, we implemented a single iteration of backward snowballing.

\begin{table}[h]
\centering
\caption{Inclusion and exclusion criteria.}
\label{tab:filter_criteria}
\resizebox{1\columnwidth}{!}{
\begin{tabular}{l}
\toprule
\multicolumn{1}{c}{\textbf{Inclusion Criteria}} \\ \midrule
\begin{tabular}[c]{@{}l@{}}1. Publications that implement software engineering methodologies, \\ approaches, and practices in test smell detection and refactoring.\\ 2. Available in digital format. \end{tabular} \\ \midrule
\multicolumn{1}{c}{\textbf{Exclusion Criteria}} \\ \midrule
\begin{tabular}[c]{@{}l@{}}1.  Publications that are not written in English.\\ 2. Websites, leaflets, and grey literature.\\ 3. Published in 2021.\\ 4. Full-text not available online. \\5. Tools not associated with peer-reviewed papers.\\  6. Duplicated publications.\end{tabular} \\ \bottomrule
\end{tabular}
}
\end{table}

To ensure the reliability of the selected studies, each study was evaluated by three authors of this paper. Each selected study underwent an agreement process, and in case of uncertainty and disagreement, we discussed it until we reached consensus. We finally ended up with with a set of 29 studies. Next, we merged the lists of test smells mentioned in these papers, which resulted in a list of 33 different test smells encountered in Java, Scala, and Android systems. The full list of papers and test smells is available online in the supplementary materials~\cite{plugin}.

Next, we considered the possibility of implementing each test smell for Python. There were several reasons why some of the test smells could not be implemented:

\textit{The test smell is not applicable to Python.} For example, the \textit{Resource Optimism}~\cite{van2001refactoring} test smell in Java occurs if a \texttt{File} object is used without checking for its existence. However, in Python, files always associate with resources, because, according to the Python official documentation, ``\texttt{open()} is the standard way to open files for reading and writing with Python''~\cite{python_files_docs}. 

\textit{The test smell detection relies on the production code that is being tested.} For example, to identify \textit{Eager Test}~\cite{van2001refactoring} and \textit{Lazy Test}~\cite{van2001refactoring}, we need to know what the corresponding production files and production classes are. A lot of recent works study test-to-code traceability~\cite{kicsi2018exploring, kicsi2020testroutes, ghafari2015automatically} and a lot of different approaches have been suggested. However, reliably making a strict one-to-one connection between a test method and a production method in the static analysis environment is difficult~\cite{kicsi2020testroutes}, which is why leave the support of such test smells for future work.

\textit{The test smell detection is possible only when the test is executed.} For example, for the \textit{Test Run War}~\cite{van2001refactoring}, it is necessary to actually run the test case, which is not possible in a static analysis environment. Even after running, identifying such test smells is non-trivial, and for practical purposes we had to exclude them.

Finally, we selected 17 test smells for implementing. We list them below.

\textbf{Assertion Roulette}
occurs when a test case has multiple non-documented assertions. Multiple assertion statements without a descriptive message impact the readability, understandability, and maintainability, as it becomes more difficult to understand the reason why this test fails~\cite{van2001refactoring}.

\textbf{Conditional Test Logic}
runs against the rule that test cases need to be simple and execute all statements in the production code. Conditions within the test case alter the behavior of the test and lead to situations where the test fails to detect defects in the production code under some conditions~\cite{peruma2019distribution}.

\textbf{Constructor Initialization}
is made by developers who are unaware of the purpose of the \texttt{setUp()} method that contains the preparation needed to perform test cases. As a result, they would define a constructor for the test suite, which is not ideal in practice~\cite{peruma2019distribution}.

\textbf{Default Test}
occurs when an IDE creates default test suites when the project is created and developers keep the default name. For example, PyCharm by default names the test suites \texttt{MyTestCase}. These suites are meant to serve as an example for developers when writing unit tests and should be renamed. Not renaming them upfront causes developers to start adding test cases into these files, making the default test suite a container of all test cases. This can also cause problems when the suites need to be renamed in the future~\cite{peruma2019distribution}. 

\textbf{Duplicate Assert}
occurs when a test case tests for the same condition multiple times~\cite{peruma2019distribution}.

\textbf{Empty Test}
occurs when a test case does not contain executable statements. Such tests are possibly created for debugging purposes and then forgotten about or contain commented out code~\cite{peruma2019distribution}. 

\textbf{Exception Handling}
occurs when passing or failing of a test case is dependent on the production method explicitly throwing an exception. Instead, developers should utilize special functionality of testing frameworks for that, such as an \texttt{assertRaises()} function~\cite{peruma2019distribution}. 

\textbf{General Fixture}
occurs when a test suite fixture is too general and some test cases only access a part of it. The fixture of a test suite is a special method that is executed before the test cases in the suite and serves as a setup step. A drawback of it being too general is that unnecessary work is being done when a test suite is run~\cite{van2001refactoring}.

\textbf{Ignored Test}
is caused by ignored test cases when it is possible to suppress some test cases from running. These ignored test cases add unnecessary overhead by increasing code complexity and making comprehension more difficult~\cite{peruma2019distribution}.

\textbf{Lack of Cohesion of Test Cases}
occurs if test cases are grouped together in one test suite but are not cohesive. Cohesion of a class is a metric that indicates how well various parts and responsibilities of a class are tied together. If test cases in a suite are not cohesive, this can cause comprehensibility and maintainability issues~\cite{greiler2013automated}.

\textbf{Magic Number Test}
occurs when assert statements in a test case contain numeric literals (\textit{i.e.}, magic numbers) as parameters instead of more descriptive constants or variables~\cite{peruma2019distribution}.

\textbf{Obscure In-Line Setup}
occurs when the test case contains too many setup steps. This can hinder inferring the actual purpose of the assertion in the test. Ideally, such preparation should be moved to a fixture or a separate method~\cite{greiler2013automated}.

\textbf{Redundant Assertion}
occurs when a test case contains assertion statements that are either always true or always false, and are therefore unnecessary~\cite{peruma2019distribution}.

\textbf{Redundant Print}
occurs when there is a print statement within the test. Print statements are considered to be redundant in unit tests as unit tests are usually executed as a part of an automated process with little to no human intervention~\cite{peruma2019distribution}.

\textbf{Sleepy Test}
occurs when developers need to pause the execution of statements in a test case for a certain duration (\textit{i.e.}, simulate an external event) and then continue with the execution. Explicitly causing a thread to sleep can lead to unexpected results as the processing time for a task can vary on different devices~\cite{peruma2019distribution}.

\textbf{Test Maverick}
was derived from the \textit{General Fixture} described above. If the test suite has a fixture with setup, but a test case in this suite does not use this setup, this test case is a maverick (outlier). The setup procedure will be executed before the test case is executed, but it is not needed~\cite{greiler2013automated}. 

\textbf{Unknown Test}
occurs when the test case has no assertion in it. It is possible to create a test case that does not use assertions, however, such a test is more difficult to understand and interpret~\cite{peruma2019distribution}.

During this selection, we also decided to focus specifically on the \textit{Unittest} testing framework~\cite{unittest} that is included into the Python Standard Library. Python also has a lot of popular third-party testing frameworks like \textit{PyTest}~\cite{pytest} and \textit{Robot}~\cite{robot}, however, certain test smells would look differently in different frameworks, and it is out of the scope of this paper to support them all. There are two reasons for choosing specifically \textit{Unittest}. Firstly, it remains one of the most popular testing frameworks in Python while also being the default one~\cite{turnquist2011python}. Secondly, according to its documentation, \textit{Unittest was originally inspired by JUnit}~\cite{unittest}, which allows us to detect some test smells from the literature that were originally proposed for \textit{JUnit}, for example, fixture-related test smells. Additionally, several other frameworks support launching test suites from \textit{Unittest}, and can therefore also be detected in this case.

\subsection{Identifying Python-specific test smells}
\label{sec:python-specific}

In addition to the test smells identified above, our goal was to include Python-specific test smells. To discover Python-specific test smells, we used a tool called \miner~\cite{golubev2021changes} to search for frequent change patterns in the histories of test suites. We explain the steps of this process in detail in this section.

\subsubsection{Project selection}
\label{sec:main_dataset}

To carry out this research, we needed to collect a dataset of mature open-source Python projects. As a starting point, we took GHTorrent~\cite{Gousi13}, a large collection of GitHub data, more specifically, their latest dump at the time of the compilation, compiled in July 2020~\cite{ghtorrent_dumps}. To process it, we used a tool called PGA-create~\cite{pga_create} that had been previously used to create Public Git Archive (PGA)~\cite{markovtsev2018public}. This tool processes the SQL dump to create a CSV file with a list of projects that facilitates their convenient filtering. Next, we selected all projects with at least 50 stars, which allowed us to filter out toy projects. We also only considered projects with Python as the main language that are not forks. This resulted in identifying 26,072 projects. Of them, we randomly selected 10,000. The reason for not simply picking top projects by stars is that testing might be organized very differently in projects of different scale, and simply picking the largest or the most popular repositories could skew our data towards a specific type of projects.

Next, we analyzed the history of the projects to find all commits where at least one Python test file was changed. We defined a Python test file as any file with the \texttt{.py} extension that has the word \texttt{test} in its filename, since Unittest has a naming convention of having the word \texttt{test} in the name of the test file~\cite{unittest}. We have conducted a small manual analysis by selecting 100 random Python files with the word \texttt{test} in their name and checking whether they are actually related to tests. In this random sample, all 100 files were related to testing, with 96 explicitly containing test suites and test cases, and another 4 containing auxiliary methods and testing utils.

4,580 of the projects had at least one commit that \textit{changed} such files. As we were looking for code changes, we selected these 4,580 projects. Since our goal was to analyze the changes themselves, for practical purposes, we decided to select a smaller set of projects using the criteria recommended in literature~\cite{kalliamvakou2014promises}. We selected projects with at least 1,000 commits, 10 contributors, 2 years since the first commit and no more than 1 year since the last push. This resulted in 450 projects. For the purposes of this paper, we will call this the \main dataset; the list is available online~\cite{plugin}.

\subsubsection{Change pattern mining}
\label{sec:miner}

To identify Python-specific test smells, we started by mining the histories of the collected projects and finding patterns in the changes made to test files that might be considered as either fixing or introducing a test smell. We extracted all changes made to Python test files from the identified 450 projects and processed these files using \miner~\cite{golubev2021changes}.

\miner is a tool that we developed for mining code change patterns in Python code. The tool is based on the algorithm developed by Nguyen et al.~\cite{nguyen2019graph} for Java. The parser in their tool is written specifically for the syntax of the Java language, and their tool stores graphs and works with them as Java objects, so we could not directly reuse the tool. At the same time, the algorithm itself is not language-specific, because it relies only on the abstract syntax trees (AST) of code before and after the change, which is why we implemented it for Python. The operation process of \miner is similar to that of the tool by Nguyen et al. Here, we briefly explain the procedure.

\miner works in two stages: building change graphs and mining patterns. In the first stage, the versions of code before and after the change are parsed into a special representation introduced by Nguyen et al. called \textit{fine-grained Program Dependence Graphs} (fgPDGs). fgPDGs are graphs with three types of nodes: \textit{data} nodes (variables, literals, constants, etc.), \textit{operation} nodes (arithmetic, bit-wise operations, etc.), and \textit{control} nodes (control sequences like \texttt{if}, \texttt{while}, \texttt{for}, etc.). These nodes are connected using two types of edges: \textit{control} edges represent a connection between a \textit{control} node and a node that it controls and \textit{data} edges show the flow of the data in the program, such edges also have labels specifying the flow of data. 

Then, unchanged nodes in the two fgPDGs of code before and after the change are connected together by special \textit{map} edges, resulting in new graphs called \textit{change graphs}. We used GumTree~\cite{DBLP:conf/kbse/FalleriMBMM14} to detect corresponding unchanged nodes in the versions before and after the change and connect them with a \textit{map} edge. This is carried out on a function level and therefore, this way, we obtain a special \textit{change graph} that represents each change to each testing function from the history of projects in our dataset. You can find an example of fgPDGs and a change graph in the supplementary materials~\cite{plugin}. 

The second stage of \miner involves searching these change graphs for patterns. This part is also done similarly to the work of Nguyen et al.~\cite{nguyen2019graph}. First, all pairs of nodes representing function calls that are also connected with the \textit{map} edge are considered to be the initial patterns that are then recursively expanded to contain new nodes. The pattern is defined by two thresholds: \textit{minimum size}, indicating the minimum number of graph nodes in the pattern, and \textit{minimum frequency}, indicating the minimum number of repetitions of the pattern in the corpus. Changing these parameters influences what is considered to be a pattern and, therefore, how many patterns are detected. This way, the patterns are expanding to detect isomorphic subgraphs within our corpus of graphs.

\begin{figure}[h]
\centering
\subfigure[Commit in the Obspy project~\cite{commit_change_obspy}.]{
	\label{subfig:correct}
	\includegraphics[height=0.27in]{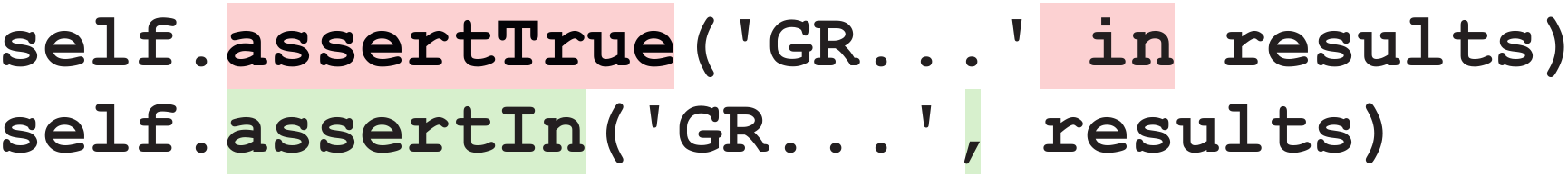}} 
	\hfill
\subfigure[Commit in the Numba project~\cite{commit_change_numba}.]{
	\label{subfig:notwhitelight}
	\includegraphics[height=0.27in]{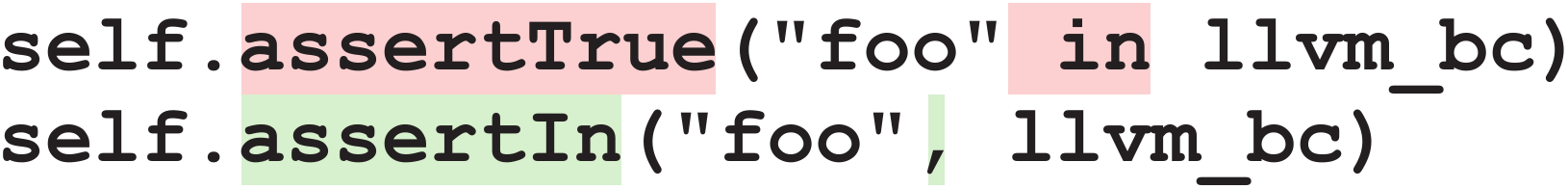}} 
	\hfill
\subfigure[Commit in the Reviewboard project~\cite{commit_better_reviewboard}.]{
	\label{subfig:nonkohler}
	\includegraphics[height=0.27in]{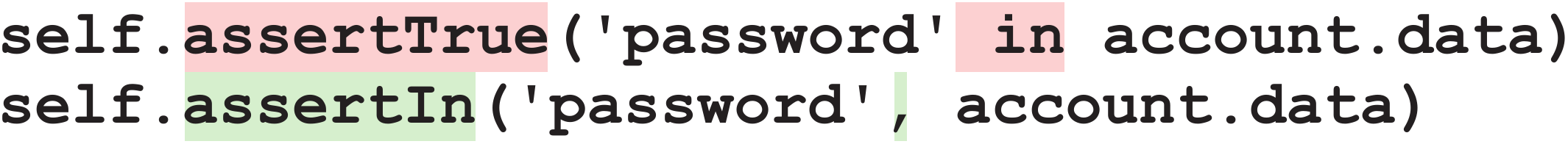}} 
	\vspace{-0.1cm}
\caption{An example of a change pattern identified in several projects on GitHub.}
\vspace{-0.1cm}
\label{fig:example_pattern}
\end{figure}

In our work, we use the same thresholds as Nguyen et al.: \textit{minimum size} of 3 and \textit{minimum frequency} of 3. It is possible that studying specifically the testing code requires different thresholds, we leave such analysis for future work. We additionally add a \textit{maximum size} threshold of 20. This is done to make the process faster by stopping the patterns from growing too large. Our own preliminary experiments and our analysis of the results of Nguyen et al. demonstrated that the majority of discovered patterns are small. More specifically, the \textit{Depth} pattern corpus provided by Nguyen et al.~\cite{nguyen2019graph} contains a total of 9,289 patterns, of which 8,697 (93.6\%) patterns are 20 nodes or smaller. Since smaller patterns are much more frequent and are easier to analyze, we decided to focus on them. An example of a discovered pattern is presented in Figure~\ref{fig:example_pattern}.

\subsubsection{Test smells detection}
\label{sec:miner_results}

In total, \miner was able to discover 8,239 different patterns in the \main dataset. Of them, 652 patterns were \textit{cross-project}, meaning they were encountered in at least two different projects, and 159 appeared in at least three different projects. Three authors of the paper independently manually labeled all 159 of such changes to discover changes that either fix or introduce possible test smells. The reason for focusing on these changes is that they are inherently more universal among different developers. Along with analyzing the code changes themselves, the authors also looked at the corresponding commit messages, since commit messages may contain the rationale for a change. After individual labeling, the authors discussed their labels and reached a perfect agreement. 

Of the studied 159 patterns, 70 (44\%) constituted various changes to assertion functionality, similar to the example shown in Figure~\ref{fig:example_pattern}. Three authors of the paper independently came to a conclusion that the candidates for possible Python-specific test smells can be found only within this group, because other common changes in testing code correlate to various other aspects of software engineering: data structures, data processing, etc., that are not directly related to testing itself. For example, popular changes include changing the level of the logger (error, info, debug, etc.) or changing the shape of a \texttt{numpy} array. Such patterns are important, but are not directly related to testing or test smells.

We categorized assert-related change patterns into three categories, which we describe below with specific examples. 

i. \textbf{Assertion changes that alter the logic.}
Often, when developers change an assertion in a test case, they do it to update the logic behind the test. For example, a pattern that occurred in six different projects is changing from \texttt{assertEqual} to \texttt{assertRegex}. This way, instead of checking for an exact equality between an object and a string, a regular expression is passed that can support variations in strings. One commit message reads: \textit{Use a more permissive comparison for jsonschema.ValidationError messages}~\cite{commit_alter_girder}. 

Another common pattern involved changing from \texttt{assertEqual} to \texttt{assertIn}, where instead of one correct result, there is a list of values. Conversely, another common example is changing from \texttt{assertIsNone} to another function \texttt{assertIsInstance}. This makes the check more specific: the object is not compared to \texttt{None} but rather needs to be an object of a new specific class. One commit message conveys a similar idea: \textit{NullSort instead of None. A more descriptive placeholder for ``don't sort''}~\cite{commit_alter_beets}.

ii. \textbf{Assertion changes that do not alter the logic and use \textit{more} appropriate functions.}

\begin{figure*}
\centering
\includegraphics[width=\textwidth]{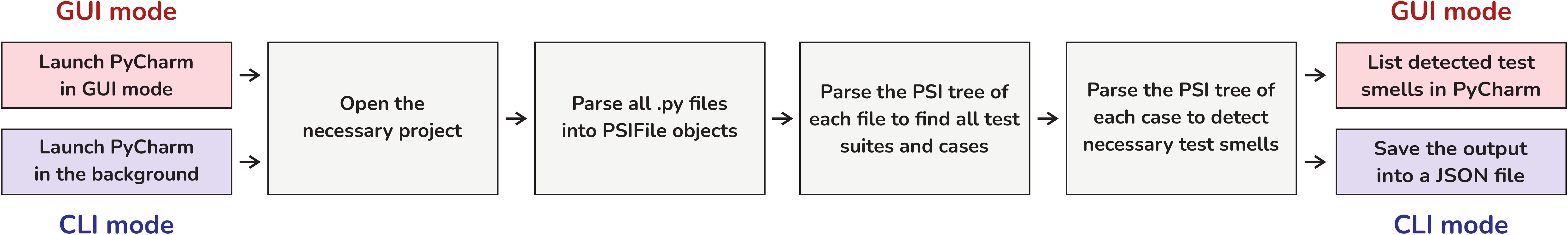}
\caption{The pipeline of \plugin operation in two regimes: GUI mode and CLI mode.}
\label{fig:plugin}
\vspace{-0.3cm}
\end{figure*}

A large portion of the patterns involved keeping the assertion logic the same, but replacing the assertion function with a more appropriate one to make the code succinct. In total, eight such patterns were identified. These changes are Python-specific in the sense that they rely heavily on a wide range of assertion functions that \textit{Unittest} supports. 

The most popular pattern is shown in Figure~\ref{fig:example_pattern}. It occurs in seven different projects and moves from using \texttt{assertTrue(X in Y)} to \texttt{assertIn(X, Y)}. One commit message describes this change in great detail: \textit{Use more specific assertions for `in' checks. A lot of old code used `assertTrue(blah in blah)', or variants on that, which didn't tell you much if there was a failure. Nowadays, we have
assertIn and assertNotIn, which we can use instead. This switches our tests to use these}~\cite{commit_better_reviewboard}. This commit message indicates that the original code (before the change) can be considered a test smell since using general assertions can make it difficult to infer the reason of failure by ``hiding'' the actual assertion in its body, whereas using specific assertions can make it easier.

Another change that strives to remove the ambiguity of a general assertion occurs in four repositories, and it moves from using \texttt{assertFalse(X == Y)} to \texttt{assertNotEqual(X, Y)}. Sometimes \texttt{assertTrue} is changed to another specific assertion. For example, in three different projects \texttt{assertTrue(X <= Y)} is changed to \textit{Unittest}'s \texttt{assertLessEqual(X, Y)}. One commit message expectedly comments this: \textit{Use more specific asserts in unit tests}~\cite{commit_better_pythonchess}. 

In Python, it is considered bad practice to check the equality of a boolean value when you can check the value itself, so in this case a boolean assertion is more correct and more interpretable, which is reflected in a common change pattern where \texttt{assertEqual(X, False)} is changed to \texttt{assertFalse(X)}.

In this section, we have given examples of some commit messages that describe the changes along with the change pattern. 
We believe that these commit messages justify considering the wrong choice of an assertion function in \textit{Unittest} as a test smell. We called this smell \smell.

iii. \textbf{Assertion changes that do not alter the logic and use \textit{less} appropriate functions.}

Interestingly, we also discovered seven change patterns that move from an appropriate assertion function to a more general one. Following the logic of the previous section, these can be treated as introducing a test smell.

The most popular such change is moving from a more specific \texttt{assertIsNotNone(X)} to a more general \texttt{assertNotEqual(X, None)}. One commit message describes this change as a \textit{Fix in test for Python 2.6 compatibility}~\cite{commit_worse_gensim}. However, the changes in this pattern were made in 2014--2015, and since Python 2 is deprecated from 2020, this is no longer a problem.

A similar message described commits in two different projects that moved from \texttt{assertNotIn(X, Y)} to \texttt{assertTrue(X not in Y)} and two more different projects that moved from \texttt{assertLess(X, Y)} to \texttt{assertTrue(X < Y)}. The same changes can be found for functions like \texttt{assertGreater(X, Y)} and \texttt{assertIsNone(X)}, with one commit message saying: \textit{remove fancy test assertions that are unavailable on 2.6}~\cite{commit_worse_requests}.

In total, we encountered 12 different suboptimal asserts (either fixed, introduced, or both). We extrapolated them to similar functions and opposite cases where necessary: for example, if there is a suboptimal assert that contains \texttt{assertLess}, it can also be formulated for \texttt{assertGreater}, etc. This resulted in a total of 32 different assertions that can be considered a part of the \smell test smell, the full list is available online~\cite{plugin}. 

\begin{table*}[t]
\footnotesize
\centering
\caption{The detection rules for supported test smells. Citations indicate works where the rules were adopted from.}
\label{table:rules}
\begin{tabular}{@{}lp{13.5cm}@{}}
\toprule
\textbf{Assertion Roulette} & A test case contains more than one assertion statement without an explanation/message.~\cite{peruma2020tsdetect}\\

\textbf{Conditional Test Logic} & A test case contains one or more control statements (\textit{i.e.}, \texttt{if}, \texttt{for}, \texttt{while}).~\cite{peruma2020tsdetect} \\

\textbf{Constructor Initialization} & A test suite contains a constructor declaration (an \texttt{\_\_init\_\_} method).~\cite{peruma2020tsdetect} \\

\textbf{Default Test} &  A test suite is called \texttt{MyTestCase}.~\cite{peruma2020tsdetect} \\

\textbf{Duplicate Assert} &  A test case contains more than one assertion statement with the same parameters.~\cite{peruma2020tsdetect}  \\

\textbf{Empty Test} &  A test case does not contain a single executable statement.~\cite{peruma2020tsdetect} \\

\textbf{Exception Handling} & A test case contains either the \texttt{try/except} statement or the \texttt{raise} statement.~\cite{peruma2020tsdetect} \\

\textbf{General Fixture} & Not all fields instantiated within the \texttt{setUp()} method of a test suite are utilized by all test cases in this test suite.~\cite{peruma2020tsdetect} \\

\textbf{Ignored Test} & A test case contains the \texttt{@unittest.skip} decorator.~\cite{peruma2020tsdetect} \\

\textbf{Lack of Cohesion of Test Cases} & The mean of the pairwise cosine similarities between test cases in a test suite $\leq$ 0.4.~\cite{palomba2018automatic}  \\

\textbf{Magic Number Test} & A test case contains an assertion statement that contains a numeric literal as an argument.~\cite{peruma2020tsdetect} \\

\textbf{Obscure In-Line Setup} & A test case contains ten or more local variables declarations.~\cite{greiler2013automated} \\

\textbf{Redundant Assertion} & A test case contains an assertion statement in which (1) the expected and actual parameters of equality are the same, \textit{e.g.}, \texttt{assertEqual(X, X)} or (2) the assertion of truth is carried out on the unchangeable object, \textit{e.g.}, \texttt{assertTrue(True)}.~\cite{peruma2020tsdetect} \\

\textbf{Redundant Print} &  A test case invokes the \texttt{print()} function.~\cite{peruma2020tsdetect} \\

\textbf{Sleepy Test} &  A test case invokes the \texttt{time.sleep()} function with no comment.~\cite{peruma2020tsdetect} \\

\textbf{Suboptimal Assert} & A test case contains at least one of the suboptimal asserts. \\

\textbf{Test Maverick} & A test suite contains at least one test case that does not use a single field from the \texttt{SetUp()} method.~\cite{greiler2013automated} \\

\textbf{Unknown Test} & A test case does not contain a single assertion statement.~\cite{peruma2020tsdetect} \\

\bottomrule
\end{tabular}
\vspace{-0.3cm}
\end{table*}

\section{\plugin}
\label{sec:tool}

Once we curated the list of test smells (explained in Section~\ref{sec:selecting}), our next goal was to implement a tool to identify them in actual Python code. We developed a tool called \plugin that currently identifies 18 test smells (17 language-agnostic from the existing literature and one Python-specific elicited by us as described in Section~\ref{sec:python-specific}), and can be run from both the graphical user interface and the command line. Figure~\ref{fig:plugin} shows the operating pipeline of \plugin. 
In this section, we explain it in greater details.

\subsection{Tool internals}
\label{sec:internals}

\plugin is implemented as a plugin for PyCharm~\cite{pycharm}, a popular IDE for Python developed by JetBrains. The plugin supports two different modes of operation: \textit{Graphical User Interface (GUI) mode} and \textit{Command Line Interface (CLI) mode}. Internally, \plugin uses Program Structure Interface (PSI)~\cite{psi} from JetBrains' IntelliJ Platform (that PyCharm is built upon) to parse Python source code and build syntactic and semantic code models for analysis. When the project is opened and the interpreter is set up, the tool uses PSI and other related PyCharm API to gather all \texttt{.py} files in the project in the form of \texttt{PSIFile} objects. 

Next, the tool extracts all Python classes that are sub-classes of \texttt{unittest.TestCase}. With the help of PSI, \plugin can deal with importing \texttt{unittest} or \texttt{unittest.TestCase} under alias or test cases that are not direct sub-classes of \texttt{unittest.TestCase}. 
After collecting individual test suites, each detector class (corresponding to each test smell) invokes \texttt{PsiElementVisitor} to create a custom visitor for the necessary \texttt{PsiElement}, which allows \plugin to identify test smells. 
For example, for the \textit{Magic Number Test}, we use a custom visitor of \texttt{PyCallExpression} to find all assertions, and then check if one of the provided arguments is a \texttt{PyNumericLiteralExpression}. If there is a match, the \textit{Magic Number Test} smell is declared to be found. 

For the test smells from the literature, we implemented their detection in the same way as they are described in the original papers, using the mentioned thresholds. For example, we detect \textit{Obscure In-Line Setup} the same way as Greiler et al.~\cite{greiler2013automated}, by counting the number of local variables in a test case and flagging the case as smelly if this number is larger than a threshold of 10, and detect \textit{Lack of Cohesion of Test Cases} the same way as Palomba et al.~\cite{palomba2018automatic}, by calculating pairwise cosine similarities between test cases. Detection rules for all the supported test smells are presented in Table~\ref{table:rules}, the citations mark the works, from where the detection rules were adapted from. Where necessary, we used code entities analogous to their counterparts in Java, for example, \texttt{@unittest.skip()} decorator in the place of the \texttt{@Ignore} annotation. If there were several different heuristics to detect the same smell in different papers, we selected one based on its recency and its convenience to implement using the PSI and the IntelliJ platfrom.

When the analysis is done, \plugin can show the detected test smells inside the IDE or save them to a JSON file for further analysis.

\subsection{Evaluation}

We conducted an experimental evaluation of the effectiveness of \plugin in correctly detecting test smells. As there are no existing datasets containing information for all the supported smells, we decided to construct our own validation set. We randomly selected eight projects that did not make it into the \main dataset. We then used the definitions of test smells to identify and tag test files with the information regarding the types of smells they exhibit. This process resulted in a total of 37 annotated files. The list of projects, together with some statistics about their testing files, is shown in Table~\ref{table:validation_projects}. To ensure an unbiased annotation process, three authors of the paper individually did the labelling and discussed their results afterwards to reach a consensus. All the three authors have experience with Python development ranging from two to five years, which includes exposure to developing unit tests.   

\begin{table}[t]
\small
\centering
\caption{Eight projects selected for the evaluation of \plugin. The columns indicate the number of testing files, suites, and cases with \textit{Unittest}.}
\label{table:validation_projects}
\begin{tabular}{@{}cccc@{}}
\toprule
\multicolumn{1}{c}{\textbf{Project}} & \textbf{T. Files} & \textbf{T. Suites} & \textbf{T. Cases} \\ \midrule
ali1234/vhs-teletext                 & 12                  & 23                  & 56                    \\
cea/sec\_ivre                        & 1                   & 1                   & 13                    \\
davidhalter/jedi                     & 3                   & 4                   & 17                    \\
demisto/content                      & 9                   & 10                  & 203                   \\
justiniso/polling                    & 1                   & 1                   & 4                     \\
Lagg/steamodd                        & 6                   & 13                  & 45                    \\
plamere/spotipy                      & 3                   & 17                  & 114                   \\
pygridtools/drmaa-python             & 2                   & 4                   & 16                    \\ \midrule
\multicolumn{1}{c}{\textbf{Total}}   & 37                  & 73                  & 468                   \\ \bottomrule
\end{tabular}
\end{table}

Next, we ran \plugin on the same set of projects and compared our results against the oracle. We calculated precision, recall, and F1 score for each test smell. We also calculated the weighted average of these three metrics for all test smells with the weights being the number of instances of each test smell in the projects. The results of the conducted evaluation are presented in Table~\ref{table:validation_results}.

Several test smells were encountered very rarely in the validation projects, with three of them having only a single example. This has to do with the fact that these test smells are just rare in Python in general (see Section~\ref{sec:test_smells_distribution}). However, these test smells have very robust definitions that are easy to detect: \textit{Default Test} requires the tool to simply check the name of the test suite, \textit{Constructor Initialization} requires the tool to simply check the presence of an \texttt{\_\_init\_\_} method, and \textit{Sleepy Test} simply looks for the \texttt{sleep()} function in the body of the test case. 

\begin{table}[t]
\small
\centering
\caption{The results of the evaluation. \textbf{Inst.} stands for instances and indicates a true number of test suites with a given smell in \\ the validation dataset.}
\label{table:validation_results}
\begin{tabular}{@{}lcccc@{}}
\toprule
\multicolumn{1}{c}{\textbf{Test Smell}} & \textbf{Inst.} & \textbf{Precision} & \textbf{Recall} & \textbf{F1} \\ \midrule
Assertion Roulette                      & 42                 & 100\%              & 97.6\%          & 98.8\%      \\
Conditional Test Logic                  & 20                 & 80\%              & 100\%           & 88.9\%       \\
Constructor Initialization              & 1                  & 100\%              & 100\%           & 100\%       \\
Default Test                            & 2                  & 100\%              & 100\%           & 100\%       \\
Duplicate Assertion                     & 6                  & 100\%              & 100\%           & 100\%       \\
Empty Test                              & 1                  & 100\%              & 100\%           & 100\%       \\
Exception Handling                      & 10                 & 100\%              & 100\%           & 100\%       \\
General Fixture                         & 11                 & 100\%              & 100\%           & 100\%       \\
Ignored Test                            & 3                  & 100\%              & 100\%           & 100\%       \\
Lack of Cohesion                        & 13                 & 78.6\%             & 84.6\%          & 81.5\%      \\
Magic Number Test                       & 23                 & 100\%              & 82.6\%          & 90.5\%      \\
Obscure Inline Setup                    & 3                  & 100\%              & 100\%           & 100\%      \\
Redundant Assertion                     & 2                  & 100\%              & 100\%           & 100\%      \\
Redundant Print                         & 2                  & 100\%              & 100\%           & 100\%       \\
Sleepy Test                             & 1                  & 100\%              & 100\%           & 100\%       \\
Suboptimal Assert                       & 10                 & 100\%              & 100\%           & 100\%       \\
Test Maverick                           & 5                  & 100\%              & 100\%           & 100\%       \\
Unknown Test                            & 10                 & 83.3\%             & 100\%           & 90.1\%      \\ \midrule
\multicolumn{1}{c}{\textbf{Weighted average}}                                 & —                  & \textbf{94.0\%}             & \textbf{95.8\%}         & \textbf{94.9\%}      \\ \bottomrule
\end{tabular}
\end{table}

As shown in Table~\ref{table:validation_results}, \plugin achieves a high level of correctness with F1 scores ranging from 81.5\% to 100\% for different test smells. For the cases where the tool did not achieve 100\%, we investigated the mismatch. 

In one instance, \textit{Assertion Roulette} was not detected because of a non-conventional name of the test case, where the name started with a \texttt{\_} symbol instead of the word \texttt{test*} as is the convention. A human rater could tag such a test case as having the \textit{Assertion Roulette} test smell, however, \plugin failed to do so. \plugin also incorrectly identified several \textit{Conditional Test Logic} test smells.  
\textit{Conditional Test Logic} is detected by the presence of control statements (\textit{i.e.}, \texttt{if}, \texttt{for}, etc.) irrespective of their impact on the assertion.
For example, the \texttt{for} statement can be used simply to assign a variable and such cases are incorrectly tagged as \textit{Conditional Test Logic} by \plugin. 
\textit{Lack of Cohesion} relies on the cohesiveness of test cases in a test suite. \plugin measures cohesiveness using cosine similarity, whereas human raters used their subjective judgement, which resulted in a mismatch between the output of \plugin and the opinion of the human raters in several cases. 
Several \textit{Magic Number Tests} were not detected because the comparison to a literal occurred in assertions with complex parameters that are not yet supported. For example, \texttt{assertEqual(df.shape, (1, ))} was tagged as a \textit{Magic Number Test} test smell by a human rater, however, \plugin failed to do so, because the literal is located in a tuple. Finally, two cases of \textit{Unknown Test} turned out to be false positives. The tool considered the test case to not have assertions, when in reality an assertion was present, but it was from the unsupported \texttt{pytest} framework.

For all test smells together, \plugin achieves the precision of 94\% and the recall of 95.8\%. Table~\ref{table:comparison} shows the comparison between the obtained values and the reported numbers of \textsc{tsDetect}~\cite{peruma2020tsdetect}, a similar tool for Java. It can be seen that the values are similar, however, we plan to conduct a more thorough and direct comparison of tools in the future.

\begin{table}[h]
\small
\centering
\caption{The comparison of performance between \plugin and \textsc{tsDetect}.}
\label{table:comparison}
\begin{tabular}{@{}lcccc@{}}
\toprule
\multicolumn{1}{c}{\textbf{Detector}} & \textbf{Language} & \textbf{Precision} & \textbf{Recall} & \textbf{F1} \\ \midrule
\textbf{\textsc{tsDetect}}~\cite{peruma2020tsdetect} & Java & 96.0\% & 97.1\% & 96.5\% \\
\textbf{\plugin} & Python & 94.0\% & 95.8\% & 94.9\% \\
\bottomrule
\end{tabular}
\end{table}

\section{Prevalence of Test Smells}
\label{sec:study}

After developing and validating \plugin, we conducted an empirical study on test smell prevalence in open-source Python projects. In this section, we present the details and the results of this study.

\begin{figure*}
\centering
\includegraphics[width=\textwidth]{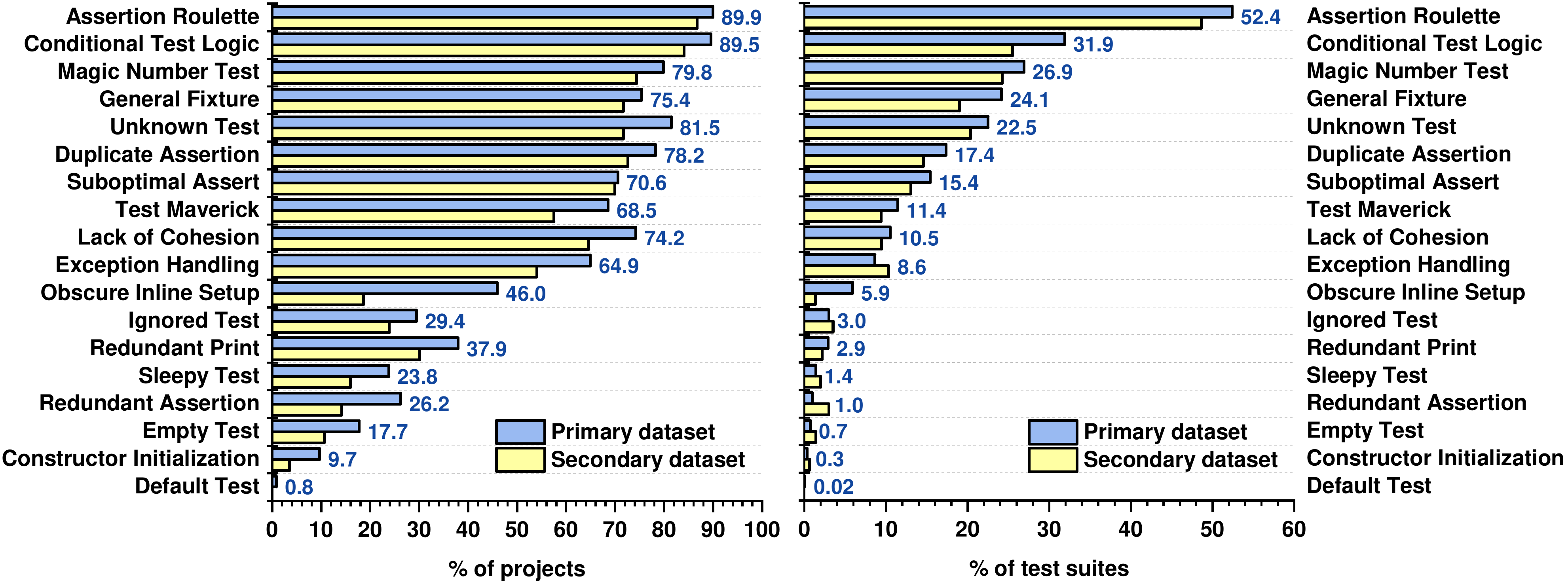}
\caption{The prevalence of different test smells among all projects and test suites in \main and \control datasets. The percentages relate to projects that use \textit{Unittest}, the numbers near bars are shown for the \main dataset.}
\label{fig:test_smells}
\vspace{-0.3cm}
\end{figure*}

\subsection{Selecting projects to analyze}

The goal of our study was to analyze test smell prevalence in Python projects using \plugin. This was done to increase the subject diversity among the existing empirical studies on test smells, as well as to gain an understanding of how test smells are diffused in Python code. We decided to study the presence of test smells in the same \main dataset that was used for mining code change patterns. We decided to do so because the \main dataset represents mature open-source Python projects that use testing within them.

However, to make sure that the results of the study are robust and do not depend on the results from Section~\ref{sec:miner_results}, we decided to also run the tool on an additional dataset. To gather it, we used the same procedure as described in Section~\ref{sec:main_dataset}, but with one condition being slightly relaxed: we gathered projects with the number of commits between 500 and 1,000, instead of at least 1,000 commits. This resulted in 239 additional projects; the full list is available online~\cite{plugin}. We will refer to this dataset as the \control dataset. While we draw our general conclusions from the \main dataset, since it contains more projects with larger histories, the purpose of the \control dataset is to make sure that the reported results are unbiased.

\subsection{Methodology}

We ran \plugin on all the projects in the \main and \control datasets separately. We dropped the results where not a single test suite was found, and only considered test suites with at least one test case and test files with at least one test suite. Test smells can occur on various levels of granularity: \textit{Constructor Initialization}, \textit{Default Test}, \textit{General Fixture}, and \textit{Lack of Cohesion} manifest at the level of a test suite as a whole, while other test smells such as \textit{Conditional Test Logic} are formulated at the test case level. 

We analyzed the test smells using their appropriate granularity. A test suite is considered smelly if it contains at least one test case with a given smell. A test file can also be considered a valid object for comparison, however, even though in Python and in \textit{Unittest} it is possible to have several test suites in one test file, this granularity is still largely similar to a test suite, and often a test file contains just one or two test suites. We also calculated the distribution of test smells among projects to get a more coarse-grained picture of the test smells prevalence.

We studied the most common and the least common test smells, as well as the prevalence of the newly proposed \smell. Additionally, we studied the co-occurrence of different test smells in individual test suites and discussed the correlations between test smells. 

\subsection{Results}

In this section, we discuss the results of the empirical study of the test smells prevalence in Python code.

\subsubsection{General information}

In total, at least one \textit{Unittest} test case was found in 248 projects out of the 450 in the \main dataset (55.1\%). From here on out, all percentages are calculated based on these 248 projects. In total, in these 248 projects, \plugin detected 9,158 test files, 16,681 test suites, and 96,736 test cases. More detailed statistics are presented in Table~\ref{table:tests}. It can be seen from the table that even mature projects vary greatly by the amount of testing within them. In our dataset, one test file on average had 1.8 test suites, and one test suite on average had 5.8 test cases. 

\begin{table}[h]
\centering
\caption{The summary of the amount of testing entities per project.}
\label{table:tests}
\begin{tabular}{ c c c c }
\toprule
 & \textbf{Test files} & \textbf{Test suites} & \textbf{Test cases}\\
\midrule
\textbf{Minimum} & 1 & 1 & 1 \\
\textbf{Mean} & 36.9 & 67.3 & 390.1 \\
\textbf{Maximum} & 323 & 870 & 5,121\\
\bottomrule
\end{tabular}
\end{table}

\subsubsection{Test smells distribution}
\label{sec:test_smells_distribution}

The distribution of 18 detected test smells is presented in Figure~\ref{fig:test_smells}. In general, it can be seen that the studied test smells are prevalent in Python code. There are only 5 projects (2\%) that have no smells, however, all of them are very small projects, with the largest having only 13 test cases. All the other projects (98\%) have tests smells in one way or another. Test smells such as
\textit{Assertion Roulette} and \textit{Conditional Test Logic} are among the most common test smells and occur in almost 90\% of projects that use \textit{Unittest} in the \main dataset. Also among the most popular test smells are \textit{Magic Number Test}, \textit{General Fixture}, and \textit{Unknown Test}. 

On the other end of the spectrum, we can see test smells that rarely occur in Python code. \textit{Empty Test} occurs in just 0.7\% of the test suites, although, interestingly, even these instances are spread out among as much as 17.7\% of the projects. \textit{Constructor Initialization} occurs in 9.7\% of the projects and 0.3\% of the test suites.
Finally, the rarest of all test smells that occurs in only two projects and only three test suites within them, is \textit{Default Test}. 
Figure~\ref{fig:test_smells} also shows that our introduced \smell smell constitutes an important addition to Python test smells: it occurs at least once in 70.6\% of the projects and 15.4\% of the test suites.

In comparison with previous works that study Java and Android code~\cite{peruma2019distribution}, it can be said that the lists of the most popular test smells generally look similar. It seems that Python code has larger percentages by projects, however, direct comparison here should be carried out in future work. One specific test smell that seems to be less prevalent in Python is \textit{Exception Handling} that occurs in 64.9\% of the projects and only in 8.6\% of the test suites. \textit{Unittest} supports a convenient list of assertions like \texttt{assertRaises}, \texttt{assertRaisesRegex} and others that may prevent the users from using \texttt{try/except} keywords in tests.

It can also be seen that the results for the \main dataset and the \control dataset are similar to each other, with the only noticeable exception being the \textit{Obscure In-Line Setup}, which is rarer in the \control dataset. The values for \smell are also similar. This demonstrates that our results obtained for the \main dataset are unbiased.

Overall, our results show that various test smells are prevalent in Python code, even some of the rarer ones still occur in more than a quarter of all projects. While some of them can be considered more subjective, others make it significantly harder to maintain the code base and to interpret the results of testing in case of failure. We hope that in the future \plugin can be used to help developers and researchers to combat the spread of test smells in their repositories.

\subsubsection{Co-occurrence of test smells}

In the previous section, we discussed how prevalent different test smells are. However, such an approach considers test smells independently from each other and does not fully describe the actual ``smelliness'' of code. To get a better understanding, we also analyzed the co-occurrence of test smells.

\begin{figure}
\centering
\includegraphics[width=\columnwidth]{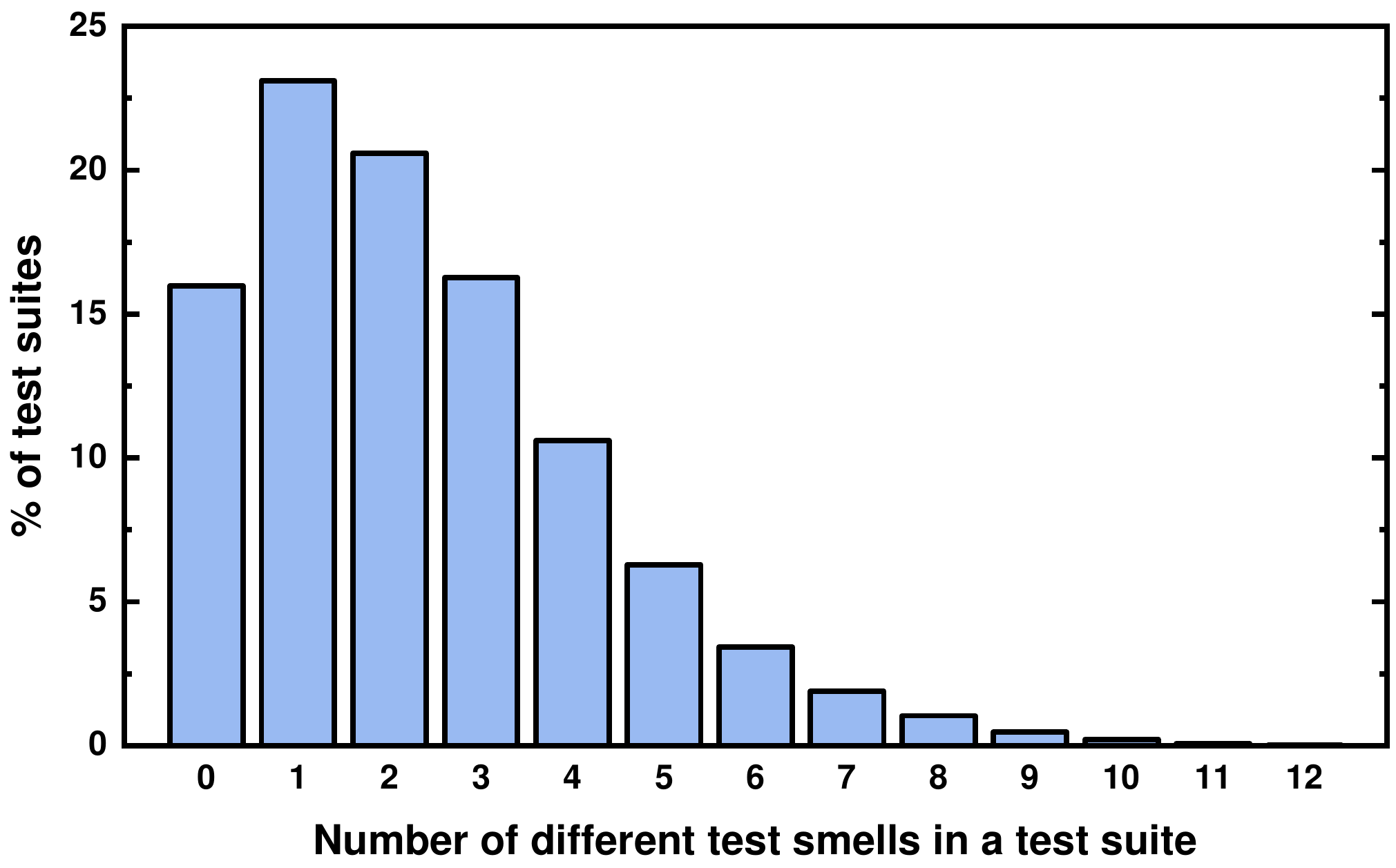}
\caption{The distribution of the number of different test smells among individual test suites.}
\label{fig:histrogram}
\end{figure}

Figure~\ref{fig:histrogram} shows the distribution of how many different smells co-exist within individual test suites. It can be seen that only 16\% of all test suites are free from smells. The remaining 84\% of the test suites have at least one smell: 23.1\% have exactly one smell, 20.6\% have two smells, 16.3\% have three smells, and this number gradually decreases with the amount of co-occurring test smells. The highest occurring number in the \main dataset appears in a single test suite with 12 distinct test smells. This large test suite with 25 test cases, in addition to all the most popular test smells, contains commented out empty test cases, catching errors with \texttt{try/except} instead of using specific assertions of error messages, and sleepy tests. It also uses \texttt{assertEqual(X, True)} instead of \texttt{assertTrue(X)}. We believe that helping developers find such suites might be useful for the maintenance of the project.

Figure~\ref{fig:histrogram} also sheds a new light on the prevalence of test smells in Python code. With more than half of all test suites having two different test smells or more, their effect on the maintainability of code can become more complex. 

We also additionally studied the co-occurrence of specific pairs of test smells. For all pairs of test smells, we calculated the following value: what percentage of test suites that have test smell X also have test smell Y. 
Two pairs of test smells are completely connected. Firstly, if the test is \textit{Empty} (\textit{i.e.}, contains no executable statements), it is automatically \textit{Unknown} (\textit{i.e.}, has no direct assertions). Secondly, if there is a \textit{Test Maverick} in a test suite, this test suite automatically has a \textit{General Fixture}. \textit{Test Maverick} occurs when the test suite has a \texttt{setUp()} method with fields and the given test case does not use any of the fields in it. Of course, this automatically means that there is at least one method that does not use all of the fields, which is the definition of a \textit{General Fixture}.
Other strongly connected pairs are all associated with \textit{Assertion Roulette} due to its popularity. If a test suite has a \textit{Duplicate Assert}, it has an \textit{Assertion Roulette} in 93.1\% of the cases. It might not be the case if the duplication has explicit messages (because \textit{Assertion Roulette} is only considered if assertions have no messages), but since it is very common to not write error messages, duplicated assertions can become a roulette. The same goes for \textit{Redundant Assertion}, 83.5\% of the test suites with which also have an \textit{Assertion Roulette}. This also makes sense, because if there is a redundant assertion, there probably should be some other assertion that is more meaningful.

This co-occurrence of test smells demonstrates that test smells have relationship with one another that should be explored in greater detail in the future. 
\section{Threats to Validity}
\label{sec:threats}

While we structured our study to avoid introducing bias and worked to eliminate the effects of random noise, it is possible that our mitigation strategies may not have been effective. This section reviews the threats to validity to our study.

It is possible that during the systematic mapping study of test smells we missed some test smells that are applicable to Python. Also, Python grammar is rather large, and is being actively updated, so \miner does not support all Python language constructs, and it is possible that we may have missed potential test smell changes because of this. We also relied on pattern detection thresholds from the original paper by Nguyen et al.~\cite{nguyen2019graph}, while it is possible that they could be different for Python and for testing code. However, the tool supports all the main features of Python and still produced a large number of code change patterns. In addition, \plugin is built in such a way that it is simple to add new test smells in the future.

The results of both parts of our study---searching for Python-specific test smells and analyzing the prevalence of test smells in Python code---rely on a specific set of open-source projects that we selected and might not generalize to all projects, including proprietary ones. However, we analyzed two moderately large datasets (\main and \control) for our tasks that were curated using various conditions suggested in the literature. We believe that the similarity of results from both datasets demonstrates the reproducibility of the results of the empirical study.

It is possible for \plugin to have some unnoticed errors in its implementation. However, we tested the tool rigorously on synthetic data and performed manual evaluation on real-world data to minimize the risk as much as possible.

One threat to validity is related to the detection of specific test smells.
Some of the implementations of test smells rely on specific thresholds that were picked from the literature. It is possible that these thresholds are different for Python, and this requires further study.


\section{Conclusions and Future Work}
\label{sec:conclusion}

Test smells are prevalent in commonly used programming languages such as Java and have a detrimental effect not only on the quality of test code but also on the production code~\cite{spadini2018relation}. 

In this work, we presented \plugin, the first tool for test smell detection in Python code that is capable of identifying 18 test smells. 17 out of these 18 test smells were adapted from test smells for other programming languages described in the literature, and we added one test smell called \smell by analyzing the most frequent changes made to test files in 450 open-source Python projects. Experiments on a set of eight real-world projects showed that \plugin is capable of detecting test smells with 94\% precision and 95.8\% recall, which is on par with other publicly available tools for test smell detection. 
Our empirical analysis shows that test smells are prevalent in Python code, with 98\% of the projects and 84\% of the test suites having at least one test smell in them.
The most frequent detected test smells were \textit{Assertion Roulette}, \textit{Conditional Test Logic}, and \textit{Magic Number Test}. 
We also observed that the proposed Python-specific \smell smell occurs in the code rather often, being present in as much as 70.6\% of the projects.

Future research directions for this work include:

\begin{itemize}
    \item Supporting more test smells, including those that rely on production code.
    \item Discovering more Python-specific smells, which requires a specific analysis of the optimal pattern searching parameters for Python.
    \item Conducting a more thorough comparison of \plugin to other tools, for example, to \textsc{tsDetect} that works with Java. It would also be of interest to employ the tools together to carry out a comparison of large Python and Java datasets from the standpoint of test smell distribution.
    \item Analyzing test smell prevalence in Python on a larger dataset of projects and in other dimensions, for example, it would be of great interest to see how test smells correlate with test coverage~\cite{virginio2019influence}.
\end{itemize}

\plugin is available on GitHub for use in the IDE and for research: \url{https://github.com/JetBrains-Research/PyNose}, all the research artifacts of this study are also publicly available: \url{https://zenodo.org/record/5156098}.

\bibliographystyle{ieeetran}
\bibliography{ase}

\end{document}